\documentclass[letterpaper,english,reprint, aps]{revtex4-1}
\usepackage[T1]{fontenc}
\usepackage[latin9]{inputenc}
\usepackage{color}
\usepackage{amsmath}
\usepackage{amssymb}
\usepackage{graphicx}

\makeatletter

\newcommand{\lyxdot}{.}

\makeatother

\usepackage{babel}
\begin{document}
\preprint{APS/123-QED}
\title{The roles of quantum coherence in thermodynamic processes}
\author{Jingyi Chen}
\affiliation{Department of Physics, Xiamen University, Xiamen 361005, Peoples Republic
of China.}
\author{Guozhen Su}
\affiliation{Department of Physics, Xiamen University, Xiamen 361005, Peoples Republic
of China.}
\author{Jincan Chen}
\affiliation{Department of Physics, Xiamen University, Xiamen 361005, Peoples Republic
of China.}
\author{Shanhe Su}
\email{sushanhe@xmu.edu.cn}

\affiliation{Department of Physics, Xiamen University, Xiamen 361005, Peoples Republic
of China.}
\author{}
\date{\today}
\begin{abstract}
Quantum coherence associated with the superpositions of two different
sets of eigenbasis vectors has been regarded as essential in thermodynamics.
It is found that coherent factors can be determined by writing observables
as an expansion in the basis vectors of the systemic density operator
and Hamiltonian. We reveal the roles of coherence in finite-time thermodynamic
processes, such as the spin precession and the spontaneous emission
of a photon. Results show that the work in the spin precession and
the heat in the spontaneous emission process are mainly generated
by coherence.

\end{abstract}
\maketitle

\section{INTRODUCTION}

The internal energy, heat, and work are fundamental quantities in
thermodynamics, regardless of states in macro or micro. The first
law of thermodynamics indicates that any change in the internal energy
of a system is given by the sum of the heat flowing into the system
and the work done by the surroundings. However, the division of the
internal energy into heat and work has become a very controversial
issue \citep{key-1-1,key-2-1,key-3-1,key-4-1,key-5-1,key-6-1,key-7-1}.
In classical thermodynamics, heat is usually defined as the flow of
energy caused by the temperature difference between the system and
its environment. On the other hand, work is performed by adjusting
the state parameters of the system \citep{key-8-1,key-9-1,key-10-1,key-11-1,key-12-1,key-13-1,key-14-1}.
More recently, quantum thermodynamics is an emerging research field
aiming to build the relation between thermodynamics and quantum mechanics
\citep{key-15-1,key-16-1,key-17-1,key-18-1,key-19-1}. One of the
important tasks is to reveal the quantum version of first law \citep{key-19-1}.
Although there have been many progresses, it is impractical to create
a unified definition of heat and work. 

For a quasi-static reversible process, the change of occupation probabilities
induces the heat transfer, while the shift of the eigenenergies of
the Hamiltonian gives rise to the work \citep{key-20-1,key-21-1}.
On the basis of the dissipative master equation, Pusz interpreted
that the work done by the external agent is associated with the time
derivative of the Hamiltonian of the working substance \citep{key-22-1}.
Alicki extended this concept and related the heat to the time derivative
of the density operator \citep{key-7-1,key-24-1}. Pusz and Alicki's
definition has been broadly applied. For example, Su et al. proposed
that quantum coherence represented by the transition between the different
quantum states contributes to heat and work \citep{key-14}.\textcolor{red}{{}
}Vu revealed the heat dissipation in connection with finite-time information
erasure and the influence of coherence in such process \citep{key-25-1}
. Cavina clarified the general bound of the efficiency of a low-dissipation
quantum Carnot engine at the maximum power \citep{key-26-1}. Tajima
established a universal framework clarifying how coherence affects
the current-dissipation trade-off relation and leads to quantum lubrication
\citep{key-27-1}. 

The application of quantum coherence has attracted interest from researchers.
In order to probably unravel the role of coherence in a concrete thermodynamic
process, it is essential to get the reliable definitions of heat and
work in advance. We will start from Pusz and Alicki's definition,
because it has been validated in Markovian master equations with static
Hamiltonian and adiabatically driven Hamiltonian\textcolor{red}{{} }\citep{key-5}.
The superpositions of two different sets of eigenbasis vectors have
been pointed out to be one of the important coherence factors \citep{key-16,key-17}.
For this reason, we are interested in addressing how this type of
coherence exerts influence on the heat and work in the finite-time
evolution of a quantum system. 

In this work, we derive the coherent term in heat and work by writing
observables as an expansion in the basis vectors of the system's density
operator and Hamiltonian. The contents are organized as follows: In
Sec. II, we introduce the quantum version of the first law of thermodynamics
for closed and open systems. In Sec. III, heat and work are written
in terms of the diagonalizing bases of the density matrix and Hamiltonian,
and their connections to coherence factors are revealed. In Sec. IV,
we demonstrate that the decomposition of heat and work is agree with
previous studies. Meanwhile, the effects of quantum coherence on the
processes of spin precession and spontaneous emission are explored.
Finally, the main conclusions are drawn. 

\section{THE FIRST LAW OF THERMODYNAMICS IN THE QUANTUM REGIME}

For an open quantum system described by the time-dependent Hamiltonian
$H$ and the density operator $\rho$, the internal energy is defined
as 

\begin{equation}
U=\operatorname{Tr}[\rho H].\label{eq:U1}
\end{equation}
During an infinitesimal thermodynamic process, the time derivative
of $U$ receives contributions from the variations of $\rho$ and
$H$,

\begin{equation}
\dot{U}=\operatorname{Tr}[\rho\dot{H}]+\operatorname{Tr}[\dot{\rho}H].\label{eq:dU1}
\end{equation}
On the other hand, the first law of thermodynamics asserts that $\dot{U}$
depend on the heat flux $\dot{Q}$ and power $\dot{W}$, i.e.,
\begin{equation}
\dot{U}=\dot{Q}+\dot{W}.
\end{equation}
Since $\dot{Q}$ and $\dot{W}$ are path functions, they will not
correspond to observables.

If the quantum system is a closed system that undergoes a thermodynamic
adiabatic process, the evolution of the density operator satisfies
the Liouville\textendash von Neumann equation \citep{key-1} 
\begin{equation}
\frac{d\rho}{dt}=-\frac{\mathrm{i}}{\hbar}[H,\rho],\label{eq:von}
\end{equation}
with $\hbar$ being the reduced Planck's constant. Equation (\ref{eq:von})
and the invariant of trace under cyclic permutations imply that $\operatorname{Tr}[\dot{\rho}H]=0$.
As the heat exchange does not exist as well, i.e., $\dot{Q}=0$, the
power performed by the external field
\begin{equation}
\dot{W}=\operatorname{Tr}[\rho\dot{H}].\label{eq:work}
\end{equation}

If no external field is applied, the Hamiltonian $H$ becomes time
independent. At this moment, an isochoric process is carried out with
the system weakly coupled to a thermal reservoir. The work done during
this process will be zero and the change in the internal energy results
from the addition or removal of heat, i.e., $\dot{Q}\neq0$ and $\dot{W}=0$.
At the same time, the first term on Eq.(\ref{eq:dU1}) is missing
and the heat flux is identified as 
\begin{equation}
\dot{Q}=\operatorname{Tr}[\dot{\rho}H].\label{eq:heat}
\end{equation}

When the system has an explicit time-dependent Hamiltonian $H$ and
is simultaneously weakly coupled to a thermal bath, the definitions
of the heat flux and power rely on the road map of the evolution of
open systems \citep{key-2}. We enumerate two situations where Eqs.
(\ref{eq:work}) and (\ref{eq:heat}) are applicable. 

For the quantum isothermal process, the Hamiltonian of the system
is adjusted slowly enough, resulting in the fact that the system remains
in equilibrium with the thermal bath at temperature $T$. In such
a quasi-static process, the heat flux and power are usually expressed
as $\dot{Q}=\sum_{n}\dot{P}_{n}E_{n}$ and $\dot{W}=\sum_{n}P_{n}\dot{E}_{n}$\textcolor{red}{{}
}\citep{key-3,key-4}, where $E_{n}$ and $P_{n}=\exp[-E_{n}/(k_{\mathrm{B}}T)]/\sum_{n}\exp[-E_{n}/(k_{\mathrm{B}}T)]$
are, respectively, the eigenenergy and the occupation probability
of eigenstate $|n\rangle$ of $H$, and $k_{\mathrm{B}}$ is the Boltzmann
constant. Therefore, the rearrangement of the occupation probabilities
induces the heat transfer between the system and the bath, while the
variation of the eigenenergy gives rise to the work performed. As
the system maintains in the state of equilibrium, the Hamiltonian
and the density operator can be written as $H=\sum_{n}E_{n}|n\rangle\langle n|$
and $\rho=\sum_{n}P_{n}|n\rangle\langle n|$. It is understandable
that the heat flux and power in the quantum isothermal process are
in line with Eqs. (\ref{eq:work}) and (\ref{eq:heat}).

If the Hamiltonian changes slowly in time, the adiabatic perturbation
theory can be applied to determine the evolution equation of the density
matrix for the open system \citep{key-5,key-6,key-7}. It has been
found that Eqs. (\ref{eq:work}) and (\ref{eq:heat}) continue to
be true for any adiabatic driving system, where the Hamiltonian associated
with the interaction with the external field must be incorporated
in $H$. 

In some cases, however, the definitions of $\dot{Q}$ and $\dot{W}$
need appropriate modifications. For the system with a weakly driven
Hamiltonian, the heat has a unique definition by applying two successive
energy measurements to the environment \citep{key-5}. For the regime
of a system subjected to a periodic driving, heat currents flow in
channels corresponding to quasi Bohr frequencies based on the Floquet-Markov
master equation \citep{key-8,key-9,key-10}. In the regime of strong
system-bath interaction, a controversial issue is how to deal with
the interaction energy and one needs to reexamine the meaning of heat
and work \citep{key-11,key-12,key-13}.

Because this work focuses on rewriting Eqs. (\ref{eq:work}) and (\ref{eq:heat})
using the representation based on the instantaneous eigenstate basis
of the density operator. The following discussion will be limited
to the conditions where the heat flux and power defined by Eqs. (\ref{eq:work})
and (\ref{eq:heat}) can be applied.

\section{HEAT AND WORK IN TERMS OF DIAGONALIZING BASES OF THE DENSITY MATRIX
AND HAMILTONIAN}

In order to reveal the roles of the coherence in thermodynamics, Eqs.
(\ref{eq:work}) and (\ref{eq:heat}) are usually written in a matrix
form with respective to a complete set of basis vectors. For the eigenket
$|n\rangle$ in the energy representation, non-zero off-diagonal element
of the density matrix $\rho_{nn^{\prime}}=\langle n|\rho|n^{\prime}\rangle$
is referred to coherence factors \citep{key-14,key-27-1}. More recently,
the factor $c_{nk}=\langle n|k\rangle$, which corresponds to the
inner product of the basis vector $|n\rangle$ of the Hamiltonian
$H$ and the basis vector $|k\rangle$ of the density operator $\rho$,
has been considered as an important factor to measure the coherence
effect \citep{key-16,key-17}. On the other hand, the time derivative
of the von Neumann entropy of the system is usually written as
\begin{equation}
\dot{S}=-k_{B}\operatorname{tr}[\dot{\rho}\log\rho]=-k_{B}\sum_{k}\dot{P}_{k}\log P_{k}\text{,}\label{eq:entropy}
\end{equation}
where $\rho=\sum_{k}P_{k}|k\rangle\langle k|$ with $P_{k}=\langle k|\rho|k\rangle$.
The entropy $S$ is closely related to the eigenvalue $P_{k}$ of
$\rho$. 

For the above reasons, it will be interesting to know how the coherence
factor $c_{nk}$ affects $\dot{Q}$ and $\dot{W}$. Rewriting Eqs.
(\ref{eq:work}) and (\ref{eq:heat}) based on the instantaneous eigenstate
bases of $\rho$ and $H$, we obtain (Appendix A)

\begin{align}
\dot{W} & =\sum_{n,k}P_{k}\dot{E}_{n}c_{kn}c_{nk}+\sum_{n,k}P_{k}E_{n}(\langle k|\dot{n}\rangle c_{nk}+H.c.)\nonumber \\
 & =\mathcal{\dot{\mathit{W}}}_{d}+\mathcal{\dot{\mathit{W}}}_{c}\label{eq:work1}
\end{align}
and

\begin{align}
\dot{Q} & =\sum_{n,k}\dot{P}_{k}E_{n}c_{kn}c_{nk}+\sum_{n,k}P_{k}E_{n}(\langle\dot{k}|n\rangle c_{nk}+H.c.)\nonumber \\
 & =\mathcal{\dot{\mathit{Q}}}_{d}+\mathcal{\dot{\mathit{Q}}}_{c},\label{eq:heat1}
\end{align}
where $H.c.$ stands for the abbreviation for the Hermitian conjugate. 

$\mathcal{\dot{\mathit{W}}}_{d}$ and $\mathcal{\dot{\mathit{W}}}_{c}$
are, respectively, the first term and the second term in the first
equality of Eq. (\ref{eq:work1}). $\mathcal{\dot{\mathit{Q}}}_{d}$
and $\mathcal{\dot{\mathit{Q}}}_{c}$ are, respectively, the first
term and the second term in the first equality of Eq. (\ref{eq:heat1}).
$\mathcal{\dot{\mathit{W}}}_{d}$ is the power corresponding to the
change of the eigenenergy $\dot{E}_{n}$, which is orignated from
the variation of the external field. $\mathcal{\dot{\mathit{Q}}}_{d}$
depends on the derivative of the distribution over time $\dot{P}_{k}$
and is accompanied by the entropy change as $\dot{S}\neq0$. $\mathcal{\dot{\mathit{W}}}_{c}$
and $\mathcal{\dot{\mathit{Q}}}_{c}$ are, respectively, caused by
the basis vectors $|n\rangle$ and $|k\rangle$. Essentially, they
are both connected with the coherence factor $c_{nk}$. The sum of
$\mathcal{\dot{\mathit{W}}}_{c}$ and $\mathcal{\dot{\mathit{Q}}}_{c}$
is equal to $\sum_{n,k}P_{k}E_{n}d(c_{nk}c_{kn})/dt$, which is exactly
the same as the change of the internal energy due to the dynamics
of coherence energy defined in Ref. \citep{key-16}. In this work,
it is clearly shown that the coherence energy may simultaneously make
contributions to the heat and work. As the discussion in Sec. II,
Eq. (\ref{eq:heat1}) could tell us that $\mathcal{\dot{\mathit{Q}}}_{d}=-\mathcal{\dot{\mathit{Q}}}_{c}$,
when the system undergoes a adiabatic process without the heat transfer.

\section{RESULTS AND DISCUSSION}

Because the density operator has two equivalent expressions $\rho=\sum_{k}P_{k}|k\rangle\langle k|=\sum_{n,n^{\prime}}\rho_{nn^{\prime}}|n\rangle\langle n^{\prime}|$,
Eqs. (\ref{eq:work1}) and (\ref{eq:heat1}) are converted into

\begin{equation}
\dot{W}=\sum_{n}\rho_{nn}\dot{E}_{n}+\sum_{n\neq n^{\prime}}\rho_{nn^{\prime}}\langle n^{\prime}|\frac{\partial H}{\partial t}|n\rangle\label{eq:work2}
\end{equation}
and
\begin{equation}
\dot{Q}=\sum_{n}\dot{\rho}_{nn}E_{n}-\sum_{n\neq n^{\prime}}\rho_{nn^{\prime}}\langle n^{\prime}|\frac{\partial H}{\partial t}|n\rangle,\label{eq:heat2}
\end{equation}
where the relations $\langle n^{\prime}|\dot{n}\rangle=-\langle\dot{n^{\prime}}|n\rangle=\langle n^{\prime}|\frac{\partial H}{\partial t}|n\rangle/(E_{n}-E_{n^{\prime}})(n\neq n^{\prime})$
have been applied. These findings give the heat flux and power accounting
the spectral decomposition of the time-dependent Hamiltonian, and
are in complete agreement with previous studies \citep{key-14,key-18}.

As an application of illustrating the effects of quantum coherence
on thermodynamic processes, we first explore the adiabatic evolution
of the spin precession in a rotating magnetic field. The Hamiltonian
of the system takes the matrix form $H_{1}=\frac{\hbar\omega_{0}}{2}\left(\begin{array}{cc}
\cos\alpha & \sin\alpha e^{i\omega t}\\
\sin\alpha e^{-i\omega t} & -\cos\alpha
\end{array}\right)$, where $\omega_{0}$ is the frequency depending on the magnitude
of the magnetic field, and the field rotates around the direction
of the Pauli-Z operator at an angular velocity $\omega$ and a tilt
angle $\alpha$. The normalized eigenvectors of $H_{1}$ are given
by $|n_{+}\rangle=\cos\frac{\alpha}{2}|e\rangle+e^{i\omega t}\sin\frac{\alpha}{2}|g\rangle$
and $|n_{-}\rangle=e^{-i\omega t}\sin\frac{\alpha}{2}|e\rangle-\cos\frac{\alpha}{2}|g\rangle$
with the orthonormal basis $|e\rangle=\left(\begin{array}{c}
1\\
0
\end{array}\right)$ and $|g\rangle=\left(\begin{array}{c}
0\\
1
\end{array}\right)$. The eigenvalue corresponding to state $|n_{\pm}\rangle$ is $E_{\pm}=\pm\frac{\hbar\omega_{0}}{2}$.
Given an initial state $\rho_{0}$, the density operator as a function
of time $\rho=U\rho_{0}U^{\dagger}$, where the evolution operator
$\begin{aligned}U & =\left(\begin{array}{cc}
U_{11} & U_{12}\\
U_{21} & U_{22}
\end{array}\right)\end{aligned}
$ with\textcolor{red}{{} }$U_{11}=(\cos\frac{\Omega t}{2}-i\frac{\omega_{0}\cos\alpha-\omega}{\Omega}\sin\frac{\Omega t}{2})e^{-\frac{i\omega t}{2}}$
, $U_{22}=(\cos\frac{\Omega t}{2}+i\frac{\omega_{0}\cos\alpha-\omega}{\Omega}\sin\frac{\Omega t}{2})e^{\frac{i\omega t}{2}}$,
$U_{12}=-i\frac{\omega_{0}\sin\alpha}{\Omega}\sin\frac{\Omega t}{2}e^{-\frac{i\omega t}{2}}$,
$U_{21}=-i\frac{\omega_{0}\sin\alpha}{\Omega}\sin\frac{\Omega t}{2}e^{\frac{i\omega t}{2}}$,
and $\Omega=\sqrt{(\omega_{0}\cos\alpha-\omega)^{2}+\omega_{0}^{2}\sin^{2}\alpha}$
\citep{key-14}. The Liouville\textendash von Neumann equation {[}Eq.
(\ref{eq:von}){]} indicates $\dot{P}_{k}=0$. From Eqs. (\ref{eq:entropy})
and (\ref{eq:heat1}), it is concluded that the unitary evolution
of a closed system leads to $\mathcal{\dot{\mathit{Q}}}_{d}=0$ and
$\mathcal{\dot{\mathit{S}}}=0$. $\mathcal{\dot{\mathit{Q}}}_{c}$
is a zero value as well, because the term inside the parentheses in
Eq. (\ref{eq:heat1}) equals zero by considering the differential
equation $i\hbar\dot{U}=HU$. Therefore, the spin carries out a thermodynamic
adiabatic evolution without the heat exchange. The detailed proofs
of $\dot{P}_{k}=0$ and $\mathcal{\dot{\mathit{Q}}}_{c}=0$ are left
for Appendix B. For the resource of the power, the partial power $\mathcal{\dot{\mathit{W}}}_{d}$
is not involved in the change of the internal energy, as the eigenvalues
$E_{\pm}$ are time independent. The power of the model is completely
generated by the coherence term $\mathcal{\dot{\mathit{W}}}_{c}=\hbar\omega_{0}\omega^{2}\sin^{2}\alpha[\cos\left(\Omega t\right)-1]/(2\Omega^{2})$.

In the next, the spontaneous emission of a photon implemented by a
two-level system interacting with the electromagnetic field is considered.
The bare Hamiltonian of the system is given by $H_{2}=E_{e}|e\rangle\langle e|+E_{g}|g\rangle\langle g|$
with $E_{e}$ being the energy level of the excited state $|e\rangle$
and $E_{g}$ being the energy level of the ground state $|g\rangle$.
Assuming that the system starts from state $\rho_{0}=\frac{1}{2}\left(\begin{array}{cc}
1 & 1\\
1 & 1
\end{array}\right)$. By using the schematic model of the amplitude-damping channel, the
matrix form of the time-dependent density operator $\rho=\frac{1}{2}\left(\begin{array}{cc}
2-e^{-\gamma t} & e^{-\gamma t/2}\\
e^{-\gamma t/2} & e^{-\gamma t}
\end{array}\right)$ with $\gamma$ being the decay rate. The detailed derivation has
been presented in Refs. \citep{key-16,key-19}. The eigenvalues of
$\rho$ are found to be $P_{e}=\frac{1}{2}e^{-\gamma t}\left(e^{\gamma t}+a\right)$
and $P_{g}=\frac{1}{2}e^{-\gamma t}(e^{\gamma t}-a)$ with $a=\sqrt{e^{2\gamma t}-e^{\gamma t}+1}$,
while their respective eigenvectors are $\left|k_{e}\right\rangle =\frac{e^{-\gamma t}(e^{\gamma t}+a-1)|g\rangle+|e\rangle}{\sqrt{e^{-\gamma t}(e^{\gamma t}+a-1)^{2}+1}}$
and $\left|k_{g}\right\rangle =\frac{e^{-\gamma t}(e^{\gamma t}-a-1)|g\rangle+|e\rangle}{\sqrt{e^{-\gamma t}(e^{\gamma t}-a-1)^{2}+1}}.$
For the spontaneous emission process, no power is generated by the
system, because the time-independent eigenvalues and eigenvectors
of Hamiltonian $H_{2}$ make the partial powers $\mathcal{\dot{\mathit{W}}}_{d}=\mathcal{\dot{\mathit{W}}}_{c}=0$
{[}Eq. (\ref{eq:work1}){]}. As a result, the work $W$ (solid line)
as a function of $t$ is a horizontal line crossing the zero point
at the vertical axis, as shown in Fig. 1. By applying Eq. (\ref{eq:heat1})
and integrating $\mathcal{\dot{\mathit{Q}}}_{d}$ and $\mathcal{\dot{\mathit{Q}}}_{c}$
over the interval $[0,t]$, Fig. 1 also presents the total heat exchange
$Q=\mathcal{\mathit{Q}}_{d}+\mathit{\mathcal{\mathit{Q}}}_{c}$ (dashed
line) as a function of $t$. The dash-double-dotted line and the dash-dotted
line represent the heat $\mathcal{Q}=\sum_{n\mathrm{,}k}\int_{0}^{t}E_{n}\left|c_{nk}\right|^{2}\frac{dP_{k}}{dt^{\prime}}dt^{\prime}$
and the work $\mathcal{W\mathit{=\sum_{n\mathrm{,}k}\int_{0}^{t}P_{k}d}}(E_{n}\left|c_{nk}\right|^{2})$
defined in Ref. \citep{key-16}, which are derived based on the classical
concept of heat and work. It is not difficult to prove that $Q=\mathcal{Q}+Q_{c}$,
$W=\mathcal{W-\mathit{Q_{c}}}$. In the spontaneous emission process,
the work performed must be zero, since no external field is applied.
The change of the internal energy of the system is only due to the
heat transfer. In Ref. \citep{key-16}, it is instructive to describe
the coherence by the factor $c_{nk}$, but the partition of heat and
work may violate the intrinsic properties of thermodynamic processes.

\begin{figure}
\includegraphics[scale=0.3]{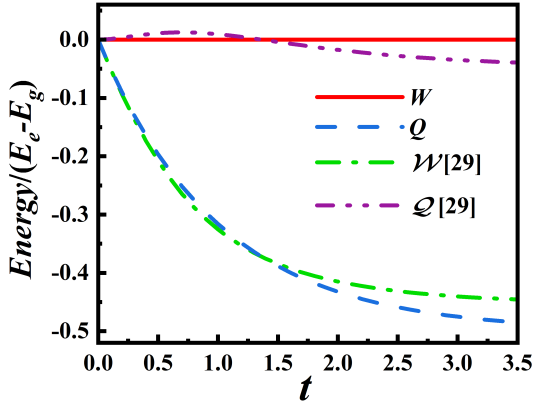}\caption{The heat $Q$ (dashed line) and $W$ (solid line) as functions of
time $t$ by integrating $\mathcal{\dot{\mathit{Q}}}$ and $\mathcal{\dot{\mathit{W}}}$
over the interval $[0,t]$. The dash-double-dotted line and dash-dotted
line are, respectively, the heat $\mathcal{Q}$ and work $\mathcal{W}$
defined in Ref. \citep{key-16} as functions of time $t$.}
\end{figure}

\section{CONCLUSIONS}

In summary, the results draw a clear distinction between the definitions
of heat flux and power in quasi-static processes and those in non-quasi-static
processes. The heat flux and power in non-quasi-static processes are
further reformulated by using the basis vectors of the density matrix
and the Hamiltonian, and their connections to the coherence are revealed.
It is demonstrated that the power is completely generated by the coherence
work in the spin precession process and the heat is mainly determined
by the coherence heat in the spontaneous emission process. This method
established here may pave the theoretical foundation for further exploring
the coherence effects in quantum thermodynamics.
\begin{acknowledgments}
This work has been supported by the National Natural Science Foundation
(Grants No. 12075197) and the Fundamental Research Fund for the Central
Universities (No. 20720210024). 
\end{acknowledgments}

\section*{APPENDIX A: THE EXPANSION OF THE HEAT FLUX AND POWER}

In the following steps, the heat flux and the power are written as
an expansion in the basis states $|k\rangle$ of $\rho$ and $|n\rangle$
of $H$:

\begin{align}
\dot{W} & =\operatorname{Tr}[\rho\dot{H}]\nonumber \\
 & =\operatorname{Tr}[\sum_{k}P_{k}|k\rangle\langle k|\frac{d\sum_{n}E_{n}|n\rangle\langle n|}{dt}]\nonumber \\
 & =\sum_{n,k,k^{\prime}}\langle k^{\prime}|(P_{k}|k\rangle\langle k|\dot{E}_{n}|n\rangle\langle n|+P_{k}|k\rangle\langle k|E_{n}|\dot{n}\rangle\langle n|\nonumber \\
 & \qquad\qquad+P_{k}|k\rangle\langle k|E_{n}|n\rangle\langle\dot{n}|)|k^{\prime}\rangle\nonumber \\
 & =\sum_{n,k}P_{k}\dot{E}_{n}c_{kn}c_{nk}+\sum_{n,k}P_{k}E_{n}(\langle k|\dot{n}\rangle c_{nk}+H.c.),
\end{align}

and

\begin{align}
\dot{Q} & =\operatorname{Tr}[\dot{\rho}H]\nonumber \\
 & =\operatorname{Tr}[\frac{d\sum_{k}P_{k}|k\rangle\langle k|}{dt}\sum_{n}E_{n}|n\rangle\langle n|]\nonumber \\
 & =\sum_{n,k,k^{\prime}}\langle k^{\prime}|(\dot{P}_{k}|k\rangle\langle k|E_{n}|n\rangle\langle n|+P_{k}|\dot{k}\rangle\langle k|E_{n}|n\rangle\langle n|\nonumber \\
 & \qquad\qquad+P_{k}|k\rangle\langle\dot{k}|E_{n}|n\rangle\langle n|)|k^{\prime}\rangle\nonumber \\
 & =\sum_{n,k,k^{\prime}}(\dot{P}_{k}E_{n}\langle k^{\prime}|k\rangle\langle k|n\rangle\langle n|k^{\prime}\rangle+P_{k}E_{n}\langle k|n\rangle\langle n|k^{\prime}\rangle\langle k^{\prime}|\dot{k}\rangle\nonumber \\
 & \qquad\qquad+P_{k}E_{n}\langle\dot{k}|n\rangle\langle n|k^{\prime}\rangle\langle k^{\prime}|k\rangle)\nonumber \\
 & =\sum_{n,k}(\dot{P}_{k}E_{n}\langle k|n\rangle\langle n|k\rangle+P_{k}E_{n}\langle k|n\rangle\langle n|\dot{k}\rangle\nonumber \\
 & \qquad\qquad+P_{k}E_{n}\langle\dot{k}|n\rangle\langle n|k\rangle)\nonumber \\
 & =\sum_{n,k}\dot{P}_{k}E_{n}c_{kn}c_{nk}+\sum_{n,k}P_{k}E_{n}(\langle\dot{k}|n\rangle c_{nk}+H.c.).
\end{align}
\textcolor{black}{Here, the unit operator $I=\sum_{k^{\prime}}|k^{\prime}\rangle\langle k^{\prime}|$
is introduced for obtaining the penultimate step of Eq. (13). Note
that the calculations of the traces with the aid of the basis states
of $H$ will obtain the same results.}

\section*{APPENDIX B:THE PROOF OF $\dot{P}_{k}=0$ AND $\dot{Q}_{c}=0$ IN
THE SPIN PRECESSION PROCESS}

In the spin precession process, the time derivative of the eigenvalue
$P_{k}$ of the density matrix $\rho$ is demonstrated to be zero
as follows
\begin{align}
\dot{P}_{k} & =\langle k|\dot{\rho}|k\rangle+\langle\dot{k}|\rho|k\rangle+\langle k|\rho|\dot{k}\rangle\nonumber \\
 & =-\frac{\mathrm{i}}{\hbar}\langle k|[H,\rho]|k\rangle+P_{k}(\langle\dot{k}|k\rangle+\langle k|\dot{k}\rangle)\nonumber \\
 & =-\frac{\mathrm{i}}{\hbar}\sum_{n,k^{\prime}}E_{n}P_{k^{\prime}}(\langle k|n\rangle\langle n|k^{\prime}\rangle\langle k^{\prime}|k\rangle\nonumber \\
 & -\langle k|k^{\prime}\rangle\langle k^{\prime}|n\rangle\langle n|k\rangle)\nonumber \\
 & =0,
\end{align}
where the Liouville\textendash von Neumann equation {[}Eq.(\ref{eq:von}){]}
and $\frac{d\langle k|k\rangle}{dt}=\langle\dot{k}|k\rangle+\langle k|\dot{k}\rangle=0$
have been applied in the second equality.

During the unitary evolution, the system changes into the state $\rho=U\rho_{0}U^{\dagger}=\sum_{k}P_{k}U|k_{0}\rangle\langle k_{0}|U^{\dagger}=\sum_{k}P_{k}|k\rangle\langle k|$,
where the diagonalized form of the initial state $\rho_{0}=\sum_{k}P_{k}|k_{0}\rangle\langle k_{0}|$.
Therefore, $|k\rangle=U|k_{0}\rangle$ can be regarded as the basis
vector $|k\rangle$ of $\rho$ at time $t$ with a constant occupation
probability $P_{k}$, and its time derivative $|\dot{k}\rangle=\dot{U}|k_{0}\rangle$.
Consequently, the heat flux $\dot{Q}_{c}$ associated with coherence
is rearranged into 

\begin{align}
\dot{Q}_{c} & =\sum_{n,k}P_{k}E_{n}(\langle\dot{k}|n\rangle c_{nk}+H.c.)\nonumber \\
 & =\sum_{n,k}P_{k}E_{n}[\langle k_{0}|\dot{U}^{\dagger}|n\rangle\langle n|U|k_{0}\rangle+H.c.]\nonumber \\
 & =\sum_{k}P_{k}[\langle k_{0}|\dot{U}^{\dagger}HU|k_{0}\rangle+\langle k_{0}|U^{\dagger}H\dot{U}|k_{0}\rangle]\nonumber \\
 & =\sum_{k}P_{k}[i\hbar\langle k_{0}|\dot{U}^{\dagger}\dot{U}|k_{0}\rangle-i\hbar\langle k_{0}|\dot{U}^{\dagger}\dot{U}|k_{0}\rangle]\nonumber \\
 & =0,
\end{align}
where the summation $\sum_{n}E_{n}|n\rangle\langle n|$ is replaced
by the Hamiltonian $H$ in the third equality, and the equation $\dot{U}=-\frac{i}{\hbar}HU$
has been applied in the fifth step.

\end{document}